\documentclass[journal=jctcce,layout=twocolumn,manuscript=article]{achemso}
\usepackage{amsmath}
\usepackage{amssymb}
\usepackage{mathtools}
\usepackage{epsfig}
\usepackage{graphicx}
\usepackage{epstopdf}
\usepackage{multicol}
\usepackage{color,soul}
\usepackage[toc,page]{appendix}

\makeatletter
\let\l@addto@macro\relax
\makeatother
\usepackage[fontsize=9pt]{scrextend}

\let\oldmaketitle\maketitle
\let\maketitle\relax

\author{Sabry G. Moustafa}
\email{smoustaf@trinity.edu}
\affiliation{Department of Engineering Science, Trinity University, San Antonio, Texas 78212, USA}
\title{Accurate and Fast Estimation of the Continuum Limit in Path Integral Simulations of Quantum Oscillators and Crystals}

\begin{document}
\twocolumn[
\begin{@twocolumnfalse}
\oldmaketitle
Convergence of path integral simulations requires a substantial number of beads when quantum effects are significant. Traditional Trotter scaling approaches estimate the continuum limit through extrapolation, however they are restricted to the asymptotic behavior near this limit. We introduce an efficient extrapolation approach for thermodynamic properties of quantum oscillators and crystals from primitive path integral simulations. The method utilizes a fitting function inspired by the analytic solution of the harmonic oscillator (HO), or the Einstein crystal for solids. The formulation is for first derivative properties, such as energy and pressure; however, extension to second derivative properties, such as elastic constants, is straightforward. We apply the method to a one-dimensional HO and anharmonic oscillator (AO), as well as a three-dimensional Lennard-Jones crystal. Configurations are sampled using path integral molecular dynamics simulations in the canonical ensemble, at a low temperature of $T=0.1$ (simulation units). Compared to Trotter extrapolation approaches, the new method demonstrates substantial accuracy in estimating the continuum limit, using only a few simulations of relatively smaller system sizes. This capability significantly reduces computational cost, providing a powerful tool to facilitate computations for more complex and challenging systems, such as molecules and real crystals. \\
\end{@twocolumnfalse}]

\section{INTRODUCTION}
Nuclear quantum effects, such as zero-point energy and tunnelling, play important roles when the thermal energy becomes smaller than the quantum energy level spacing. This is the case, in general, at low temperatures and for systems containing light nuclei. With dense phases and molecule, specifically, the extent of ``quantumness'' is also controlled by the stiffness of the force constants. Imaginary-time path integral simulation provides a powerful and robust tool to numerically account for these effects. In this framework, the quantum system is mapped onto an extended classical isomorphism, made of $n$ replicas (also referred to as beads or the Trotter number) of the original system. These replicas are connected consecutively using harmonic springs, in a closed ring-polymer arrangement of $n$ beads. This provides an approximate description, however, the true quantum system is recovered in the continuum limit ($n\to\infty$).

The ``primitive'' approximation (PA) represents the ``standard'' discretization method for path integrals, which is based on the Trotter factorization scheme. In this formulation, the approximate partition function $Z_n$ of a system of $N$ distinguishable particles at a temperature $T$, occupying $d$-dimensional space, is given by the following integral over a $dNn$ extended phase space,\cite{tuckerman2023book}
\begin{subequations}
\label{eq:ZV_PI}
\begin{align}
\label{eq:Z_PI}
Z_n\left(\beta \right) &= \left(\frac{m n}{2\pi \hbar^2 \beta}\right)^{dNn/2} \int {\rm d} {\bf x} 
\; {\rm exp}\left(-\beta V\left({\bf x}, \beta\right)\right), \\
\label{eq:V_PI}
V \left({\bf x}, \beta\right) &=  \frac{1}{n}\sum_{i=0}^{n-1} \frac{1}{2} m \omega_n^2\left({\bf x}_i - {\bf x}_{i-1} \right)^2 + \frac{1}{n} \sum_{i=0}^{n-1}  U\left({\bf x}_i\right),
\end{align}    
\end{subequations}
where $m$ is the atom mass, $\beta=1/k_{\rm B}T$, with $k_{\rm B}$ the Boltzmann constant, $\hbar= h/2\pi$ is the reduced Planck constant, $\omega_n = n/\beta\hbar$, and $\bf x$ is a coordinates vector of all the $Nn$ beads, and  ${\bf x}_{i}$ is the coordinates of the $i^{\rm th}$ replica (of length $dN$). Here, $V$ represents a $T$-dependent effective potential, which comprises the intermolecular potential $U\left({\bf x}_i\right)$ from all replicas and and the kinetic energy, represented by harmonic interactions of identical spring constants, $m\omega_n^2/n$. 

It is well-known that the leading term of this partition function, and its associated properties, is of order $\left(\beta/n\right)^2$ in the asymptotic limit.\cite{suzuki1985general} Hence, the number of beads required for statistical convergence is proportional to $1/T$, such that $nT$ remains constant. The value of the proportionality constant depends on the target statistical uncertainty (set by the simulation length)  --- larger values are needed with precise data. With molecules and crystals, in particular, the convergence is achieved with $n$ values given by multiples of $\beta\hbar\omega_{\rm max}$ (``quantumness''\cite{ceriotti2018review}), where $\omega_{\rm max}$ is the maximum vibrational frequency (Debye frequency for crystals).\cite{herrero2014path,ceriotti2018review}  Although this method yields converged results in a single run, it may not feasible to perform when a large number of beads is required (e.g., at low temperatures) or when using computationally expensive models (e.g., DFT).

Higher order approximations, such as the fourth-order Takahashi-Imada action\cite{takahashi1984tia}, can, in principle, provide accurate estimates using relatively smaller system sizes. However, they rely on higher derivatives of the intermolecular potential (e.g., Hessian), which are not always available. An alternative approach involves running simulations on relatively smaller system sizes, using the PA formulation, and then extrapolating the data to the continuum limit.\cite{brualla2004higher,chin2010extrapolated,sakkos2009high} As long as the extrapolation estimates are accurate, the standard PA method should be sufficient.

A common fitting function is given by the asymptotic (quadratic) behavior of the PA method. However, this is merely a special case of a more general series expansion of a property $Q_n$, in even powers of $\tau\equiv\beta/n$ (imaginary-time increment)
\begin{align}
\label{eq:trotter_scaling}
Q_n = Q_{\infty} + c_2 \tau^2 + c_4 \tau^4 + \dots ,
\end{align}
where, $Q_{\infty}$ is the desired continuum limit, and $c_i$ are constants.  This procedure is often referred to as Trotter scaling (or extrapolation).\cite{muser1995path} Note that, since the series is expanded around $\tau=0$, it is necessary to include relatively large system sizes (small $\tau$) to ensure accurate extrapolation. On the other hand, the extent of how small system sizes can be involved in the fit is determined by the order of the polynomial used. Typically, second (asymptotic limit)\cite{brualla2004,blinov2004path,hinsen2012path}, fourth,\cite{suzuki1985general,knoll2000molecular,bogojevic2005,martovnak1998,schmidt2018} or higher order\cite{mielke2015improved} polynomials are employed.

Moreover, Suzuki introduced an expression to also describe the asymptotic behaviour, but using a closed-form,\cite{suzuki1985general,suzuki1986quantum}
\begin{align}
\label{eq:suzuki_func}
Q_n=Q_{\infty} + \frac{a\tau^2}{1 + b\tau^2}  ,  
\end{align}
where $Q_{\infty}$, $a$ and $b$ are constants, to be determined through fitting. It can be realized that expanding this function around $\tau=0$ reduces to the Trotter scaling form (Eq.~\ref{eq:trotter_scaling}). To the best of the Author's knowledge, this expression has not been used for extrapolation purposes before this work.

Both of these scaling forms are applicable in general, regardless of the thermodynamic state. However, for quantum oscillators and crystals, in particular, the analytically solvable (for any $n$) harmonic approximation offers an opportunity for improvement. For example, Cuccoli et al.\cite{cuccoli1992quantum} leveraged this fact to compute a harmonic finite-size correction to thermodynamic properties (e.g., total energy and heat capacity). Although their results showed smaller finite-size effects compared to raw data, the extrapolated value is still approximate as it neglects the finite-size effects in the anharmonic contribution. Hence, the accuracy of the extrapolated estimate can not be systematically improved in such an \textit{ad hoc} approach. 

In this work, we also utilize the harmonic approximation, but to directly fit the raw data of quantum oscillators and crystalline systems. The fitting form is inspired by the exact solution of the harmonic oscillator (HO), or the Einstein crystal for solids, which is known at finite Trotter numbers. The formulation is designed for thermodynamic properties based on first derivatives of the free energy, however we focus on the total energy as an example. We consider 1D HO and anharmonic oscillator (AO), as well as a 3D Lennard-Jones (LJ) crystal, at a relatively low temperature ($T=0.1$) where quantum effects are substantial. The method shows high accuracy in describing finite size effects to a substantially small $n$, relative to both the Trotter (Eq.~\ref{eq:trotter_scaling}) and Suzuki (Eq.~\ref{eq:suzuki_func}) scaling forms. Accordingly, one can accurately estimate the continuum limit in these systems using few simulations of relatively small system sizes ($n$). This, in turn, offers a fast (yet, accurate) procedure to simulate quantum systems at low temperatures, which is otherwise computationally demanding, or even not feasible. While we only consider two simple models, the method would be especially useful when working with real molecules and solids using expensive first-principles methods.

\section{METHODS}
\subsection{Harmonic Oscillator}
\label{sec:methods_ho}
Inspired by the exact solution of the HO, or the Einstein crystal for solids, we formulate a new fitting function to describe finite size effects in the thermodynamic properties of quantum oscillators and crystals. The potential energy of a one-dimensional HO is given by
\begin{align}
\label{eq:U_ho_1d}
U_{\rm HO}\left(x\right) &= \frac{1}{2} m\omega^2 x^2,
\end{align}
where $m$ and $\omega$ are the mass and angular frequency of the oscillator (or the Einstein crystal), respectively, and $x$ is its position. The associated finite-$n$ partition function and Helmholtz free energy in the PA discretization method (Eq.~\ref{eq:ZV_PI}) are given, respectively, as\cite{schweizer1981,cao1993born,chin2023analytical} 
\begin{subequations}
\label{eq:ZAn_ho}
\begin{align}
\label{eq:Zn_ho}
Z_n^{\rm HO} &= \frac{1}{2\sinh{\left(\frac{n\alpha}{2}\right)}}, \\
\label{eq:An_ho}
\beta A_n^{\rm HO} &=\ln\left(2\sinh\left(\frac{n\alpha}{2}\right)\right),
\end{align}
\end{subequations}
where $\alpha \equiv 2\sinh^{-1}\left(\frac{\epsilon}{2}\right)$ and $\epsilon \equiv \beta \hbar  \omega/n$. Replacing $n$ with $\beta\hbar\omega/\epsilon$, we can recognize that both quantities depend explicitly on the system size through $\epsilon$, along with an explicit dependence on the quantumness, $\beta\hbar\omega$. In the continuum limit, we get $n\alpha\to\beta\hbar\omega$ and, hence, Eqs.~\ref{eq:ZAn_ho} reduce to the well-known HO results; namely, $Z_{\infty}^{\rm HO}=1/\left(2\sinh\left(\frac{\beta\hbar\omega}{2}\right)\right)$ and $\beta A_{\infty}^{\rm HO}=\ln\left(2\sinh\left(\frac{\beta\hbar\omega}{2}\right)\right)$.

Thermodynamic properties are given as derivatives of the free energy with respect to some external perturbation $\nu$. Here, we focus on first derivative properties, which are expressed for the HO model as:
\begin{align}
\label{eq:Qn}
Q_n^{\rm HO} &\equiv \frac{\partial \beta A_n^{\rm HO}}{\partial\nu}
= \frac{\hbar}{2}\frac{\partial \beta \omega }{\partial \nu}\frac{\coth\left(\beta\hbar\omega \frac{\alpha}{2\epsilon} \right)}{\sqrt{1+\frac{1}{4}\epsilon^2}} ,
\end{align}
where we used $n =\beta \hbar  \omega/\epsilon$, along with the chain rule for the $\alpha\left(\epsilon\right)$ derivative. It can be shown that expanding this expression, around $\epsilon=0$, yields polynomials in even powers of $\epsilon$, which is consistent with the Trotter scaling (Eq.~\ref{eq:trotter_scaling}). Note that, in this ideal model, the dependence on the Trotter number is universal, i.e., it does not depend on the specific property considered. In this work, we consider the total energy $E_n$ as an example; however, the analysis also hold for other first derivative properties (e.g., pressure). Replacing $\nu$ with $\beta$ in Eq.~\ref{eq:Qn} yields
\begin{align}
\label{eq:En_ho}
E_n^{\rm HO}\left(\beta\hbar\omega\right)= \frac{\beta\hbar\omega}{2} \frac{\coth\left(\beta\hbar\omega\frac{\sinh^{-1}\left(\frac{\epsilon}{2}\right)}{\epsilon} \right)}{\sqrt{1+\frac{1}{4}\epsilon^2}}  , 
\end{align}
While the square root function depends solely on the system size (through $\epsilon$), the hyperbolic cotangent function depends also on the quantumness level $\beta\hbar\omega$. Nevertheless, due to the properties of the $\coth$ function, its variation with $\epsilon$ is less as $\beta\hbar\omega$ increases. In this study, we are interested in highly quantum systems ($\beta\hbar\omega\gg 1$), such that most variations are described accurately by the square root function.

\subsection{Fitting Function}
In order to describe finite-size effects in real oscillators and crystals, a relaxed form to Eq.~\ref{eq:En_ho} is needed. However, the form must reduce back to the universal Trotter scaling (Eq.~\ref{eq:trotter_scaling}) when expanded around $\epsilon=0$. Here, we propose the following four-parameter function, which satisfies this condition,
\begin{align}
\label{eq:fit}
E_n\left(\epsilon\right) &= \frac{a}{\left(1+b \epsilon^2+d\epsilon^4\right)^c},
\end{align}
where $a=E_{\infty}$, $b$, $c$ and $d$ are fitting parameters. Applying this form to HO data would then yield $b\approx \frac{1}{4}$, $c\approx \frac{1}{2}$, and $d\approx 0$ (see Eq.~\ref{eq:En_ho}). However, deviations from these values are expected when applied to non-harmonic systems.

We compare the fitting accuracy of this form against a first (Poly1) and second (Poly2) order expansions (in $\epsilon^2$) of the Trotter scaling (Eq.~\ref{eq:trotter_scaling}), and the Suzuki form (Eq.~\ref{eq:suzuki_func}). All the fits were performed using the weighted least squares method, with $1/\sigma_i^2$ as the weights, where $\sigma_i$ is the uncertainty in the data (error bar). We used the \texttt{curve\_fit} routine from SciPy package to perform the fits.\cite{virtanen2020scipy} The fitting script reports both the parameters and their associated uncertainties.

\subsection{Centroid Virial Estimator}
We adopt the centroid virial estimator, due to its known computational advantages over others  (e.g., thermodynamic or virial estimators).\cite{shiga2005cv} For example, the statistical uncertainty in a measured property is known to be independent of the number of beads.\cite{herman1982path} In this method, the average total energy $E_n$ of a $d$-dimensional system of $N$ species and $n$ beads, is given by
\begin{subequations}
\begin{align}
 \label{eq:E_cvir}
E_n  &= \frac{dN}{2\beta} + \left<  \sum_{i=0}^{n-1} \frac{1}{n}  U\left({\bf x}_i\right)   - \frac{1}{2}  \sum_{i=0}^{n-1}  {\bf F}_i^{0} \cdot   \left( {\bf x}_i - {\bf x}_c\right)  \right>,
\end{align}   
\end{subequations}
where ${\bf x}_c \equiv \frac{1}{n}\sum_{i=0}^{n-1} {\bf x}_i$ is a vector (length $dN$) of the centroid coordinates associated with each atom  and ${\bf F}^{0}_i\equiv-\frac{1}{n}\frac{\partial U\left({\bf x}_i\right)}{\partial {\bf x}_i}$ is the intermolecular force vector (length $dN$) associated with the $i$ replica.

\subsection{Computational Details}
Three models are considered in this study: one-dimensional quantum HO and AO models, and a three-dimensional LJ FCC crystal of $N$ particles. For the AO model, we adopt the following quadratic-quartic form,
\begin{align}
\label{eq:U_ao_1d}
U_{\rm AO}\left(x\right) &=\frac{1}{2}m\omega^2 x^2 + k_4 x^4,        
\end{align}
where $m\omega^2$ and $k_4$ are the force constants associated with the quadratic (harmonic) and quartic terms, respectively. For both the HO and AO models, we set $m=1$, $\omega=1$, $\hbar=1$, and $k_4=0.1$. The associated quantumness degree is then $\beta\hbar\omega=10$, at the given temperature ($T=0.1$).

For the FCC crystal, we adopt the standard 12-6 LJ model. The potential is truncated and force-shifted to ensure zero forces at the cutoff distance ($r_{\rm c}$),
\begin{align}
U_{\rm LJ}\left(r\right) &= u\left(r\right) - u\left(r_{\rm c}\right) 
-\left(r-r_c\right) u'\left(r_{\rm c}\right),
\end{align}
where $u\left(r\right) \equiv 4 \epsilon_{\rm LJ} \left[ \left(\frac{\sigma_{\rm LJ}}{r}\right)^{12} - \left(\frac{\sigma_{\rm LJ}}{r}\right)^{6}\right]$ and $u'$ is its first derivative, $r$ is the pair separation. The energy and distance in this model are given in the reduced units of the LJ size ($\sigma_{\rm LJ}$) and energy ($\epsilon_{\rm LJ}$) parameters, which are set to unity throughout this work. We set the Boltzmann constant to $k_{\rm B}=1$ in these units, for both the oscillators and the LJ models. The system is composed of $N=256$ particles, at a number density of $\rho=1.0$, arranged on a $4\times4\times4$ supercell of FCC unit cells. Results did not show (statistical) sensitivity to using bigger system sizes. We set a truncation radius of $r_{\rm c}=3.0$, which is less than half the box size. The angular frequency associated with the Einstein crystal model is computed according to $\omega=\sqrt{\frac{k_2}{m}}$, where $k_2$ is the self term of the Hessian force constants of the LJ crystal, which is found to be $k_2=218.211$. We use a LJ mass of $m=1$, hence the associated frequency is $\omega=14.772$. Two values of $\hbar$ are considered, $0.1$ and $0.2$, which correspond to a quantumness levels of $\beta\hbar\omega=14.772$ and $29.544$, respectively, at the given temperature ($T=0.1$). Note that, according to Eq.~\ref{eq:ZV_PI}, the path integral properties are functions of the combined parameter $\frac{\hbar^2}{m}$ (sometimes used as inherent quantumness\cite{wiebe2020phase}). Accordingly, fixing $\hbar$ and using a different mass (while fixing $\frac{\hbar^2}{m}$) would yield the same quantumness and, hence, same results. However, for convenience, we instead choose to fix the mass in order to be able use the same time step size in the molecular dynamics simulations at different quantumness levels. 

We use path integral molecular dynamics (PIMD) simulation to sample the configurations in NVT ensemble. The HO/AO simulations ran for $10^7$ steps, after $10^6$ steps of equilibration, with a step size of $\Delta t=0.1$; whereas, the LJ simulations ran for $10^4$ steps, after a $10^3$ steps of equilibration, using $\Delta t=0.01$. For efficient sampling, the recently introduced ``HO staging'' coordinates are used to propagate the coordinates.\cite{moustafa2024staging} The method uses $\omega$ as an input, for which we use the aforementioned values for the HO, AO, and LJ systems. Moreover, to ensure canonical NVT ensemble, we adopted the white noise Langevin (stochastic) thermostat, using the BAOAB integrator method.\cite{Leimkuhler2013,liu2016simple,moustafa2024staging} The thermostat friction coefficient was set to $\gamma=\omega$, which is the optimum value in the absence of external field\cite{liu2016simple}. All the simulations were performed using the Etomica simulation software,\cite{schultz2015etomica} which can be accessed at https://github.com/etomica/etomica/tree/path\_integral

For the given level of precision, we found that using a number of beads of $n=20\beta\hbar\omega$ yields statistically converged results, regardless of the model. This corresponds to $\epsilon=0.05$, which we use as the smallest value for the finite-size analysis. This corresponds to $n=200$ for both the HO and AO models, and to $n=296$ and $593$ for the low and high quantumness levels considered, respectively. The other values are determined from fixed increments of $0.2$ in $\epsilon^2$. We then take the nearest integer of the $\beta\hbar\omega/\epsilon$ value as the corresponding number of beads. Note that this process could yield the same $n$, which we then skip to the next $\epsilon^2$ step.

The statistical uncertainties (error bars) in ensemble averages are estimated from single runs using the block averaging technique (100 blocks). \cite{book2023understanding}  All error bars reported correspond to 68\% confidence limits (i.e., $\pm \sigma$). We notice that uncertainties in raw data are statistically independent of $n$ (for $n \gtrapprox 3$), which is well-established observation with the centroid virial estimator, as mentioned earlier. Hence, this homoscedasticity property results in similar weights for all data points in the weighted least square procedure.

\section{RESULTS AND DISCUSSION}
In this section we first present the dependence of the raw data of the HO, AO, and LJ systems on the Trotter number. We then compare extrapolation results from our proposed fitting function, and compare the accuracy against traditional fitting approaches (Trotter scaling and Suzuki expression).

\subsection{Finite Trotter Number Effects}
\begin{figure}
\centering
\includegraphics[width=0.45\textwidth]{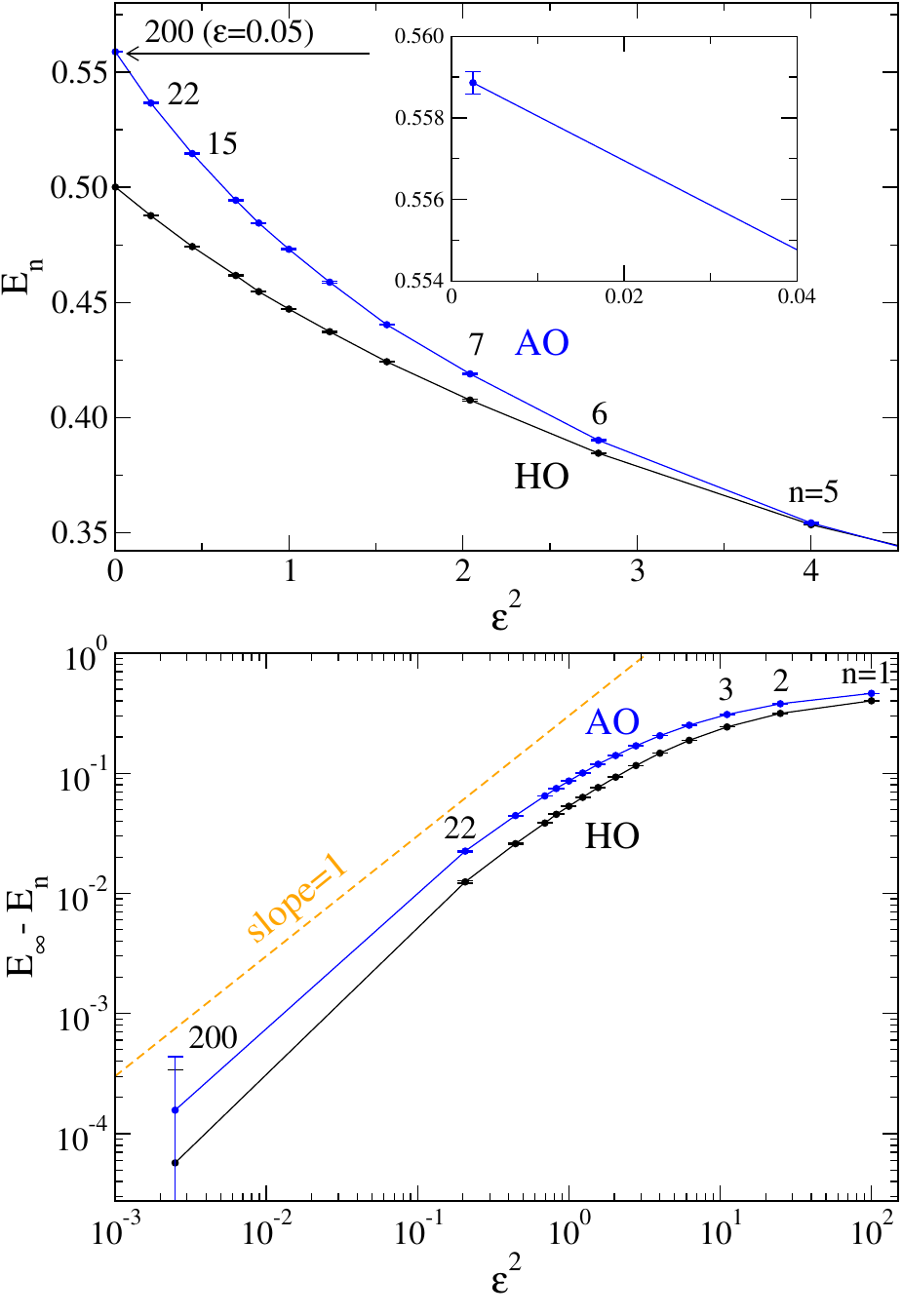}
\caption{Finite Trotter number effects on the total energy of the HO and AO models, at $T=0.1$. The corresponding quantumness is $\beta\hbar\omega=10$. (Top) absolute value; (bottom) difference from the continuum limit on a log-log scale. Lines simply join the point. The data are presented as a function of $\epsilon^2$, the natural independent variable to describe variations ($\epsilon=\beta\hbar\omega/n$, hence, $\epsilon=10/n$.). The smallest $\epsilon$ is 0.05 for both models, which corresponds to $n=20\beta\hbar\omega$, the minimum $n$ needed to get statistically converged results (see the inset). The dashed line ($y=a\epsilon^2$) is included to help identify the power of the leading term. Error bars, here and throughout this work, correspond to a $68\%$ confidence limits.}
\label{fig:fse_ao}
\end{figure}

Figure~\ref{fig:fse_ao} (top) shows the dependence of the total energy of the HO and AO models on the Trotter number. The temperature is chosen to be low enough ($T=0.1$) in order to have a challenging state of high quantumness ($\beta\hbar\omega=10.0$). However, similar results were obtained using higher temperatures (not shown). It is evident that both models have monotonic variations with respect to $\epsilon^2$, which is the natural independent variable in the PA approximation (see Eq.~\ref{eq:trotter_scaling}). However, the AO model shows a quantitatively different finite-size effects, apparently due to anharmonic effects. Apparently, the effect of anharmonicity is not merely a constant shift of the HO results. Therefore, at least for this model, using the harmonic correction approximation\cite{cuccoli1992quantum} with the AO model would not yield an accurate extrapolation.  

Moreover, in order to the investigate how the data approach the continuum limit ($E_{\infty}$), we present in Fig.~\ref{fig:fse_ao} (bottom) the difference from that limit, which represents the systematic error. The exact limit was used with the HO model, $E_{\infty}=\frac{\hbar\omega}{2}\coth\left(\frac{\beta\hbar\omega}{2}\right)$, whereas for the AO case, we estimated the limit using a linear extrapolation based on the three leftmost points. The data are presented on a log-log scale in order to detect the power of the leading term. As one would expect from the Trotter scaling (Eq.~\ref{eq:trotter_scaling}), the leading term in both cases is of order ${\cal O}\left(\epsilon^2\right)$, which is determined by comparing the slope to the dashed line. The only difference in behavior between both models is the proportionality constants of the leading terms. Therefore, even in that near-continuum limit, the harmonic correction\cite{cuccoli1992quantum} would not provide a reliable extrapolation. On the other hand, the quadratic behavior appears to be reached (statistically) only at very small $\epsilon$ values (around two leftmost points). Accordingly, using this asymptotic behaviour for extrapolation would still require a large number of beads, which makes the computational cost comparable to running a single simulation with $n=20\beta\hbar\omega$ beads (see the inset of the top subplot). 

\begin{figure}
\centering
\includegraphics[width=0.45\textwidth]{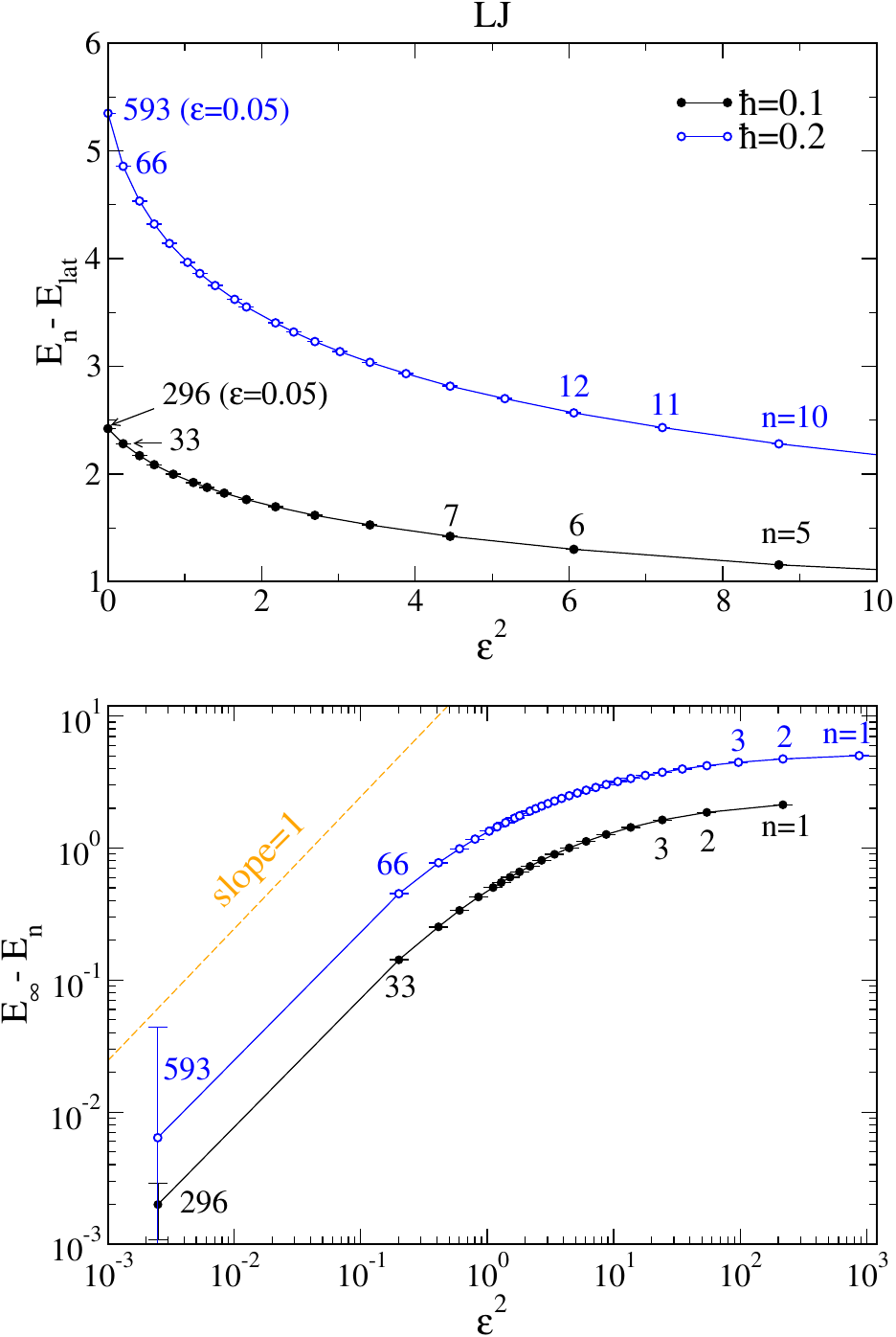}
\caption{Same as Fig.~\ref{fig:fse_ao}, but for the LJ crystal at $T=0.1$ and a density of $\rho=1.0$. The data are presented as the difference from the lattice contribution (top) and the continuum limit (bottom), respectively. Two quantumness values are considered $\beta\hbar\omega=14.772$ and $29.544$, corresponding to $\hbar=0.1$ and $0.2$, respectively.}
\label{fig:fse_lj}
\end{figure}

Figure~\ref{fig:fse_lj} shows a similar type of data, but for the total energy of the LJ crystal. The top subplot shows the thermal contribution to the energy; i.e., after subtracting the $n$-independent lattice energy ($E_{\rm lat}=-7.321$). This is the quantity we need to fit using the proposed fitting function, as shown below. This is not the case with the traditional fitting approaches, which use the absolute values. We consider both $\hbar=0.1$ and $0.2$, corresponding to the lower ($\beta\hbar\omega=14.772$) and higher ($\beta\hbar\omega=29.544$) quantumness levels investigated here. Similar to the HO and AO cases, $\epsilon=0.05$ (first point) yields statistical convergence, which corresponds to $n=296$ and $593$ for the low and high quantumness levels, respectively.  A first observation to report here is that the finite-size effects depend on the quantumness degree, such that finite-size analysis are needed for each case. Same conclusion holds for the case of HO and AO models (not shown). Moreover, the difference data from the bottom panel suggest a leading term of order ${\cal O}\left(\epsilon^2\right)$, again, consistent with the Trotter scaling of the PA approximation. Although both quantumness levels yield same leading asymptotes, their proportionality constants are different. A similar conclusion to the HO/AO models can be drawn here regarding the inadequacy of the quadratic asymptote to extrapolate to the continuum limit using relatively small system sizes. It is worth pointing out that similar results (not shown) are observed with the case thermal pressure, which could be attributed to the fact that both quantities are first derivatives as mentioned earlier (see Sec.\ref{sec:methods_ho}).

\subsection{Continuum Limit: $E_{\infty}$}

\begin{figure}
\centering
\includegraphics[width=0.45\textwidth]{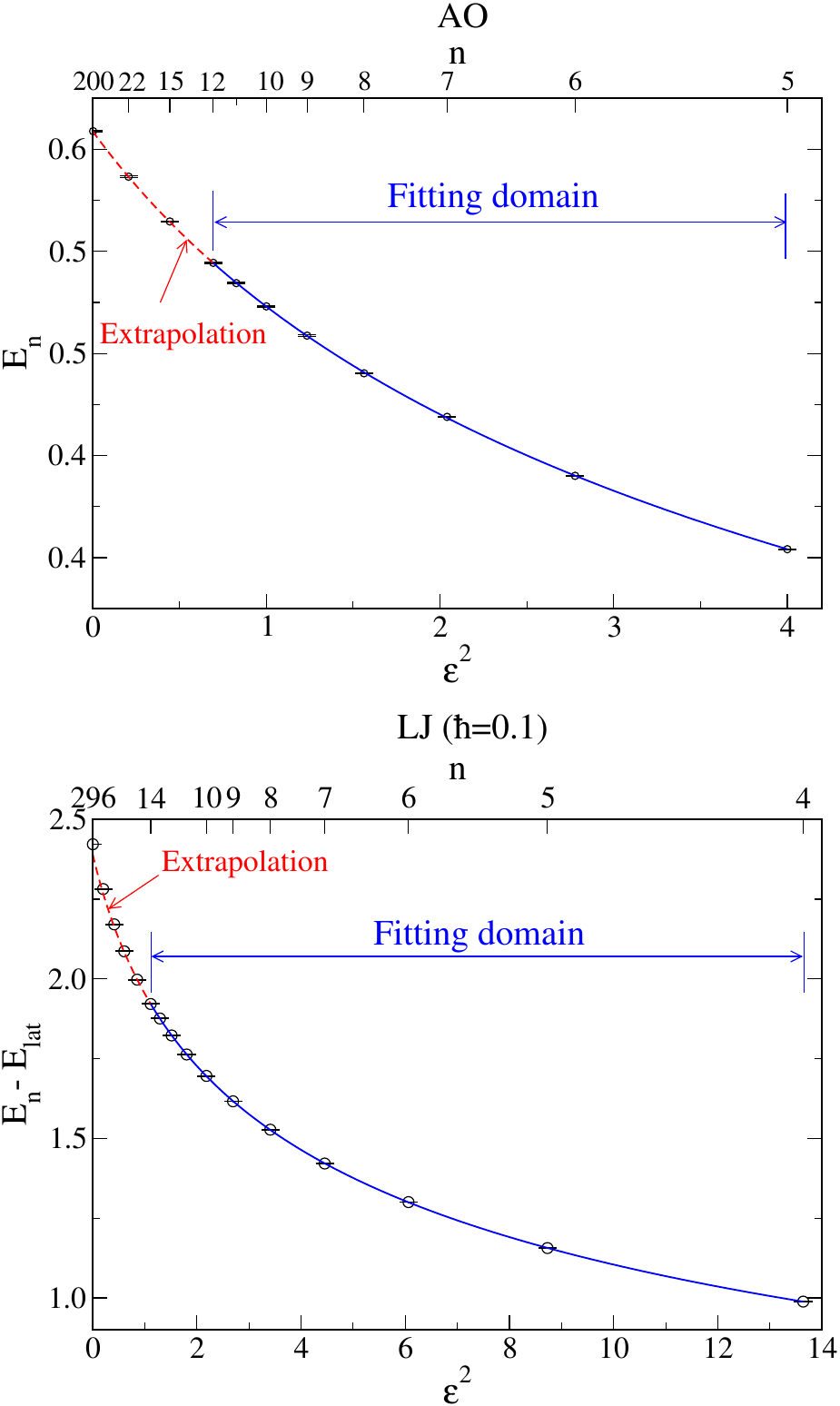}
\caption{Fitting results using Eq.~\ref{eq:fit} for the AO and LJ ($\hbar=0.1$) data, with the fitting domains comprise 8 and 11 points, respectively. The corresponding $n$ values are shown in the top $x$-axis. The red dashed line is an extrapolation of the fits to the continuum limit.}
\label{fig:fits}
\end{figure}

We estimate the continuum limit by, first, fitting the raw $E_n$ data versus $\epsilon^2$ (top panels of Figs.~\ref{fig:fse_ao} and~\ref{fig:fse_lj}), using Eq.~\ref{eq:fit}. Then, we use the value and uncertainty of the fitting parameter $a$ as an estimation to $E_{\infty}$ and associated uncertainty. Although we already have data for system sizes large enough to yield statistically converged results ($\epsilon=0.05$), we perform the fits to only a subset of data, far from the continuum limit. This allows us to investigate the fit accuracy, by comparing extrapolated estimates against direct measurements at $\epsilon=0.05$. Moreover, this would be the procedure one would follow with expensive models, such as \textit{ab initio} models. Figure~\ref{fig:fits} shows example fits for the AO and LJ ($\hbar=0.1$) data. In each case, the fitting domain is chosen such that the reduced $\chi^2$ parameter (measures the goodness of the fit) is around 1.0. This correspond to 8 and 11 data points (simulations) for the case of AO and LJ ($\hbar=0.1$) models, respectively, with the corresponding number of beads shown in the top $x$-axis. The effect of the starting point (rightmost point of the fitting domain) is discussed next.

\begin{figure}
\centering
\includegraphics[width=0.45\textwidth]{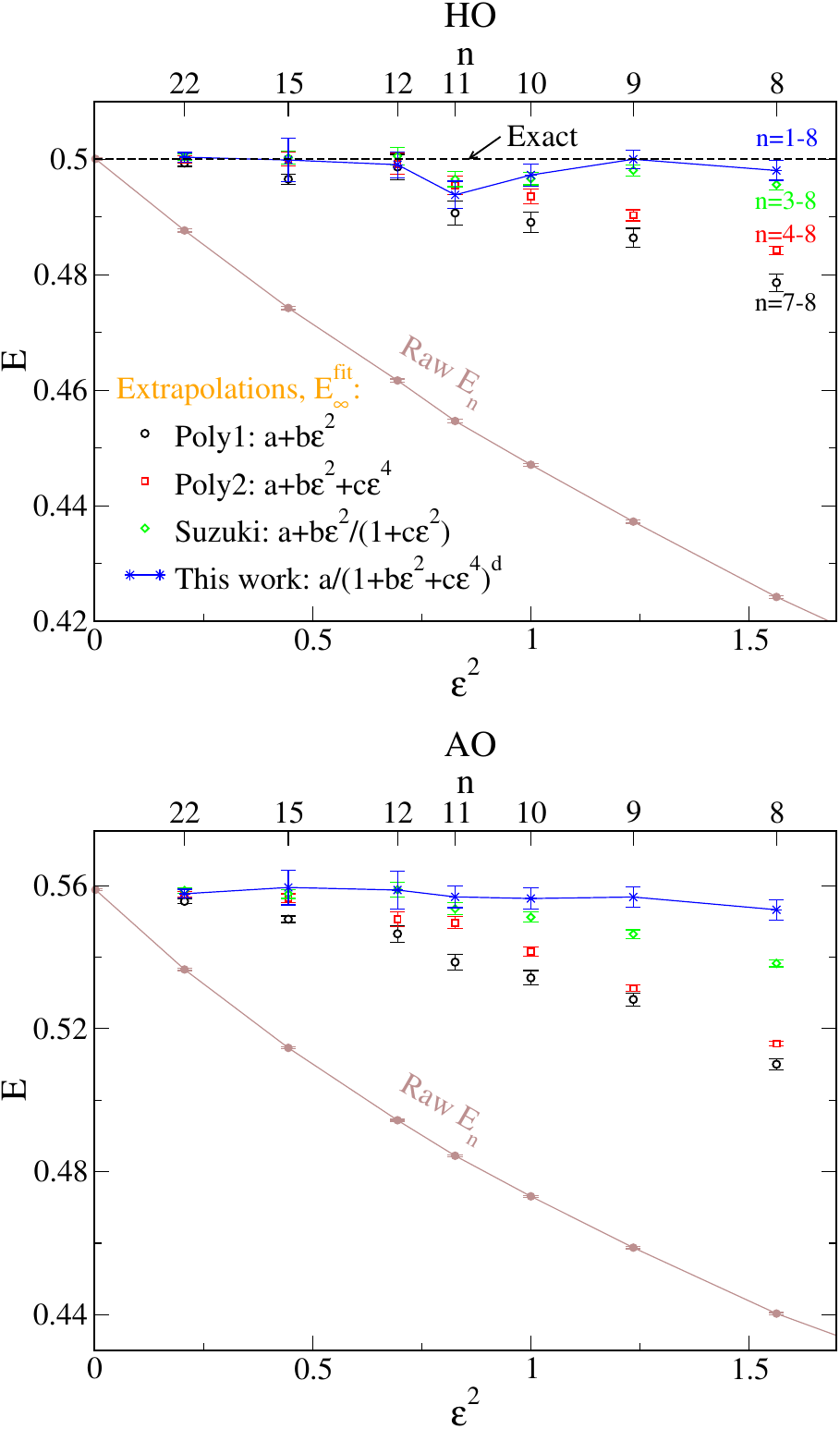}
\caption{Continuum limit estimates of the energy ($E_{\infty}^{\rm fit}$) of the HO (top) and AO (bottom) models, using different fitting functions, along with the raw data (grey line). The fitting domains include 8 data points for the HO and AO models. The $x$-axis for the extrapolated estimates represents the leftmost $\epsilon^2$ value of the fitting domain, with the corresponding $n$ value shown in the top $x$-axis. The ranges of the $n$ values for the first point (from the right) is indicated as an example.  The exact energy for the HO is given by $E_{\infty}=\frac{\hbar\omega}{2}\coth\left(\frac{\beta\hbar\omega}{2}\right)$.}
\label{fig:fit_inf_ao}
\end{figure}

Now, we investigate the accuracy of the continuum limit estimates using different fitting forms. We report results using the traditional Trotter scaling (Eq.~\ref{eq:trotter_scaling}, second and fourth orders) and the Suzuki closed-form (Eq.~\ref{eq:suzuki_func}) against the proposed expression (Eq.~\ref{eq:fit}). Figure~\ref{fig:fit_inf_ao} shows extrapolation results for the energy ($E_{\infty}^{\rm fit}$) of the HO (top) and AO (bottom) models, along with the raw data (grey). The $x$-axis associated with the four extrapolated estimates represents the leftmost $\epsilon^2$ point of the fitting domain, with the corresponding $n$ value shown in the top $x$-axis. The error bars represent the uncertainty in the fitting parameter associated with $E_{\infty}^{\rm fit}$. As expected, the accuracy of the fitting forms is the highest as the starting point of the fitting domain approaches the continuum limit. However, it is evident that the accuracy of the Trotter scaling polynomials is the lowest, while the proposed form provide the most accurate estimates. On the other hand, the performance of the Suzuki form falls somewhere in between the Trotter scaling and current fitting function. According to the data, running only 8 simulations ( $n=1,2,\dots,8$) of a total number of beads of $36$, and fit using the new form, would yield an estimate to the continuum limit with accuracy comparable to that from a single run with $n=200$ beads. This, in turn, results in a considerable reduction of the computational resources needed. 

\begin{figure}
\centering
\includegraphics[width=0.45\textwidth]{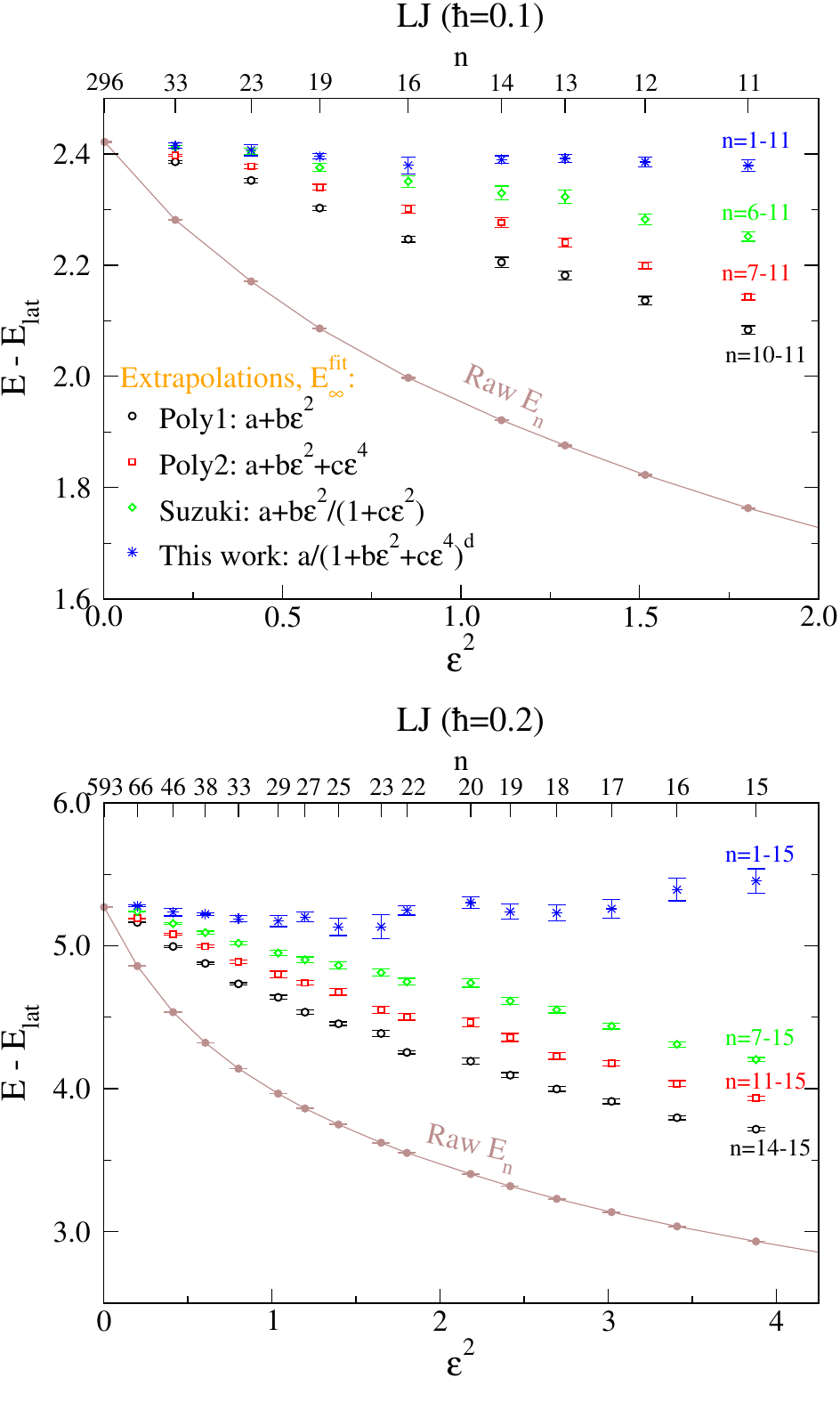}
\caption{Same as Fig.~\ref{fig:fit_inf_ao}, but for the LJ crystal at $\beta\hbar\omega=29.544$ (left, $\hbar=0.2$) and $14.772$ (right, $\hbar=0.1$). The fitting is performed on the data given in Fig.~\ref{fig:fse_lj} (top raw), for the thermal energy (top) and \textbf{pressure} (bottom). The fits start from the smallest $\epsilon=0.05$ ($n=296$ for $\hbar=0.1$ and $n=593$ for $\hbar=0.2$) to an upper value corresponds to $n$ beads given by the (inverted) $x$-axis.}
\label{fig:fit_inf_lj}
\end{figure}

Figure~\ref{fig:fit_inf_lj} shows the same type of analysis, but for the LJ crystal, using the data in Fig.~\ref{fig:fse_lj}. The performance of the fitting functions, at both the quantumness levels, is qualitatively similar to that of the oscillators. However, the continuum limit estimates using the new form appear to be less accurate with the higher quantumness case. This is to be expected because the $\hbar=0.2$ case requires double of the number of beads compared to the $\hbar=0.1$ case (see the different $x$-axis ranges).  It is interesting that the proposed form still provides an accurate description with the crystalline system, even though they were formulated based on a simple HO model. Of course, actual crystals have a wide range of normal mode vibrations, rather than a single one. However, the quality of these fits suggests that the simple Einstein crystal model (a collection of identical and independent HOs) can provide a good description of the finite-size effects for crystals.

\section{CONCLUSIONS}
We have introduced an efficient and accurate approach to estimate the continuum limit of the energy of quantum oscillators and crystals using the primitive path integral simulations. The method involves fitting data from only a few simulations of a relatively small system sizes (beads) and extrapolating to the continuum limit. This enables fast and accurate estimates, even for highly quantum states (e.g., at low temperatures) and/or expensive models, which could be otherwise computationally challenging.

The proposed fitting function used for extrapolation is based on the exact analytical solution of the harmonic oscillator but includes additional flexibility to account for anharmonic effects. To investigate its performance, we apply this approach to a one-dimensional HO and AO, and a three-dimensional LJ crystal, at a sufficiently low temperature ($T=0.1$) where quantum effects are substantial. For the purpose of comparison, we also include extrapolation results from traditional Trotter scaling and Suzuki methods, which are based on the asymptotic behavior near the continuum limit. 

In all cases considered, the new fitting form shows remarkable accuracy in predicting the continuum limit of the energy, in comparison to traditional extrapolations. This includes accurate estimates from considerably small system sizes. Therefore, a reliable estimation of the continuum limit can be achieved with much less computational effort compared to running a converged simulation with many beads ($n=20\beta\hbar\omega$) or simulating near the asymptotic limit and extrapolating.

The efficiency and simplicity of our approach provide a robust, fast, and accurate tool for estimating thermodynamic properties using the path integral technique. This is especially useful at low temperatures, where quantum effects are significant. Although we focused on energy as an example, the formulation is applicable to other first-derivative properties (e.g., pressure). Additionally, extending the analysis to second-derivative properties, such as heat capacity and elastic constants, should be straightforward. While we implemented this approach on simple models as a proof of concept, it is readily applicable to more complex models, such as molecules and real crystals.

\begin{acknowledgement}
The author thanks Dr. Andrew J. Schultz and Prof. David A. Kofke for insightful discussions.
\end{acknowledgement}

\bibliography{main}
\end{document}